# Conformal Nanocarbon Coating of Alumina Nanocrystals for Biosensing and Bioimaging

*Morteza Aramesh, [1,2,3,4]1 Phong A. Tran, [2] Kostya (Ken) Ostrikov, [2,3] and Steven Prawer [4]*

1. *Laboratory of Biosensors and Bioelectronics, Institute for Biomedical Engineering, ETH Zürich, Zürich 8092, Switzerland*
2. *School of Physics, Chemistry and Mechanical Engineering, Institute for Health and Biomedical Innovations, and Institute for Future Environments, Queensland University of Technology (QUT), Brisbane QLD 4000, Australia*
3. *CSIRO-QUT Joint Sustainable Processes and Devices Laboratory, Lindfield NSW 2070, Australia*
4. *School of Physics, The University of Melbourne, Melbourne, Victoria 3010, Australia*

**Abstract**

A conformal coating technique with nanocarbon was developed to enhance the surface properties of alumina nanoparticles for bio-applications. The ultra-thin carbon layer induces new surface properties such as water dispersion, cytocompatibility and tuneable surface chemistry, while maintaining the optical properties of the core particle. The possibility of using these particles as agents for DNA sensing was demonstrated in a competitive assay. Additionally, the inherent fluorescence of the core alumina particles provided a unique platform for localization and monitoring of living organisms, allowing simultaneous cell monitoring and intra-cellular sensing. Nanoparticles were able to carry genes to the cells and release them in an environment where specific biomarkers were present.

* Corresponding author: E-mail: mrtz.aramesh@gmail.com; maramesh@ethz.ch (Morteza Aramesh)

Carbon nanomaterials are emerging as promising material platform in nano-medicine and nano-biotechnology [1]. Nanocarbon materials appear in different forms and dimensions such as quantum dots, nanocrystals, nanotubes, 2D graphene layers, nanoporous membranes and other 3D nanoarchitectures [2]. The broad diversity in chemical and physical properties of carbon is unique compared to any other element on the periodic table, opening a wide range of opportunities for applications such as electrocatalytic intracellular sensors, neural electrodes, bionic devices and drug/gene delivery agents [3]. However, for some practical applications, carbon needs to be implemented with other materials to provide multifunctional properties [2]. Hybrid composite nanocarbon structures can potentially offer more diverse properties which cannot be obtained with carbon alone. Therefore, development of hybrid nanocarbon composites can promise new possibilities for advanced applications, such as magnetic resonance imaging, photothermal/radio-frequency therapy, multimodal cellular imaging and gene delivery [4].

Over the last decade, fluorescent nanoparticles have become key component of many important biological applications such as biosensing, bioimaging, drug delivery and drug discovery [5]. An appropriate nanoparticle-system as an agent for bioapplications should exhibit multifunctionalities such as tunable surface and optical properties, chemical stability and biocompatibility [6]. Many different types of nanoparticle systems are being introduced to resolve some of the current issues in cell-based studies [7]. Alumina nanoparticles with their natural abundancy and also strong photoluminescence in the near-infrared (NIR) region (~ 690-710 nm within the tissue transparency window) have some potentials for being implemented in biological assays [8]. However, direct application of alumina nanoparticles in biosystems is hindered due to surface properties (limited water dispersibility, chemical stability and biocompatibility). Carbon coating of alumina nanoparticles could potentially resolve many of those limitations while maintaining the optical properties. However, the conformal coating of the particles faces several considerable technical challenges due to limitations in chemical and physical properties [9].

Here we introduce a fabrication method which allows conformal coating of alumina nanoparticles with an ultra-thin carbon layer using a plasma-assisted technique. We investigate the properties of these materials and the formation mechanism by transmission microscopy (TEM), Raman spectroscopy and X-ray photoemission spectroscopy (XPS). A proof-of-principle study was carried out to show that the fabricated core-shell particles can be used as a promising platform for *in vitro* biosensing and bioimaging. In particular, the particles were able to quench the fluorescence of molecular dyes and were successfully implemented in a competitive assay for DNA sensing. Also, the internalization of the particles into the cells was investigated allowing live cell imaging and intra-cellular sensing. The results show that the fabricated core–shell nanoparticles can be used as selective and sensitive signalling platform for observation and eventually engineering biological processes at the subcellular level.

The coating method is based on the exposure of alumina particles to hydrogen/methane plasmas, in which a layer of amorphous carbon was grown on the surface of the particles (supporting information for methods and growth mechanism). **Figure 1** summarizes the approach to fabricate carbon coated alumina particles. The approach is universal for alumina particles with different shapes and sizes (from nm to mm size). Microscope images show that the originally white alumina particles turned to black colour after the plasma induced reaction. TEM images from the cross-sections of larger particles reveal the atomic structure of core-shell material and its comparison to the non-coated particles. The results show the ultra-thin conformal carbon layer with a thickness of 2-5 nm around the particles. It also suggests that the crystalline structure of the core alumina (α-alumina) remains unchanged after the plasma treatment.

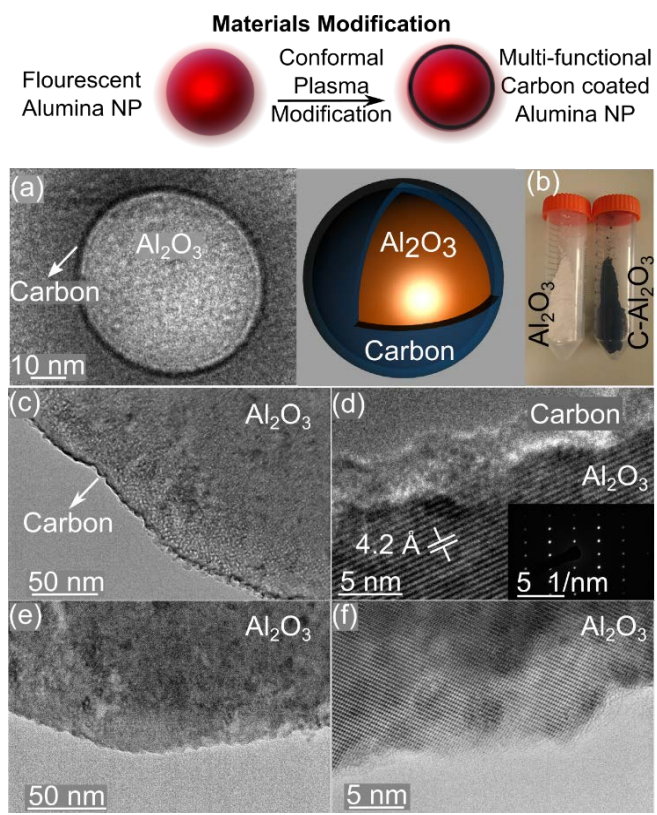

*Figure 1*. Conformal carbon coating of alumina particles with plasma. (a) TEM image and schematic of a carbon coated alumina nanoparticle. (b) A photograph of alumina particles before and after carbon coating (~35 grams in weight). Carbon coating of the particles also transforms the color of the alumina particles from white to black (c) TEM image from focused-ion-beam milling cross-sectioned carbon coated particle (see SI). The dark line around the edges of the particle is the thin carbon layer. (d) High resolution TEM image from the interface of the core alumina and carbon layer. The inset shows the selected area (electron) diffraction (SEAD) of the core alumina with its crystal structure. The lattice spacing is measured to be 4.2 Å which corresponds to the D spacing of (1120) planes in α-$Al_2O_3$. (e,f) TEM images of uncoated alumina particles. In comparison to the coated particles, the thin amorphous carbon layer is not present in these images.

The hybridization state of the carbon layer can be controlled during the growth by changing the plasma power. Different growth temperature (650–1300 °C) was achieved by varying the input power from 1250 W to 3000 W in the plasma discharge. The Raman spectrum of the samples grown at different temperatures are shown in **Figure 2**. Raman studies suggest that the influence of the temperature on the hybridization state of the grown nanocarbon layer was significant. The spectrum of the carbon coated particles contains two significant modes at 1347 cm$^{-1}$ (D peak) and 1607 cm$^{-1}$ (G peak), which are associated with the modes of amorphous carbon (a-C) films [10]. Pristine alumina particles did not show any detectable resonance in this region of Raman spectrum. The intensity ratio between D and G peaks (*I(D)/I(G)*) varied with temperature (**Figure 2(b,c)**). A higher ratio was obtained for the films that were grown at lower temperatures (650-800 °C). Also, as confirmed by XPS (high-resolution C 1s spectra), sp$^3$ content of the films decreased with temperature (**Figure 3(a)**). The grown film became considerably graphitic-like at 900-1100 °C (major sp$^2$ content with disordered structure). At temperatures higher than 1100 °C, coatings exhibited more ordered sp$^2$ bonding, as was observed in their narrow Raman peak at 2691 cm$^{-1}$. The amorphous carbon layer transforms to graphene at temperatures as high as 1300 °C.

It is particularly evident by comparing the relative carbon content obtained by XPS spectra (**Figure 3(b,c)**) before and after coating, that the intensity of the carbon signal was reduced significantly after the plasma-induced coating process. A high amount of carbon was observed in ground alumina particles, most probably due to incorporation of different hydrocarbons and organic molecules during the sample preparation process. The carbon content consistently decreases with annealing temperature by getting consumed in different forms such as gas production and volatilization during densification of the alumina matrix at the interface [11]. In a carbon-rich plasma, however, diffusing carbon atoms can also get involved in chemical interactions at the solid-gas interface (see SI). These interactions possibly produce different types of carbide and oxycarbides at the interface (supported by observation of Al–C peak in the Al 2p XPS spectrum in **Figure 3(d)**), and subsequently a network of the carbon atoms forms on the outer surface of the particles.

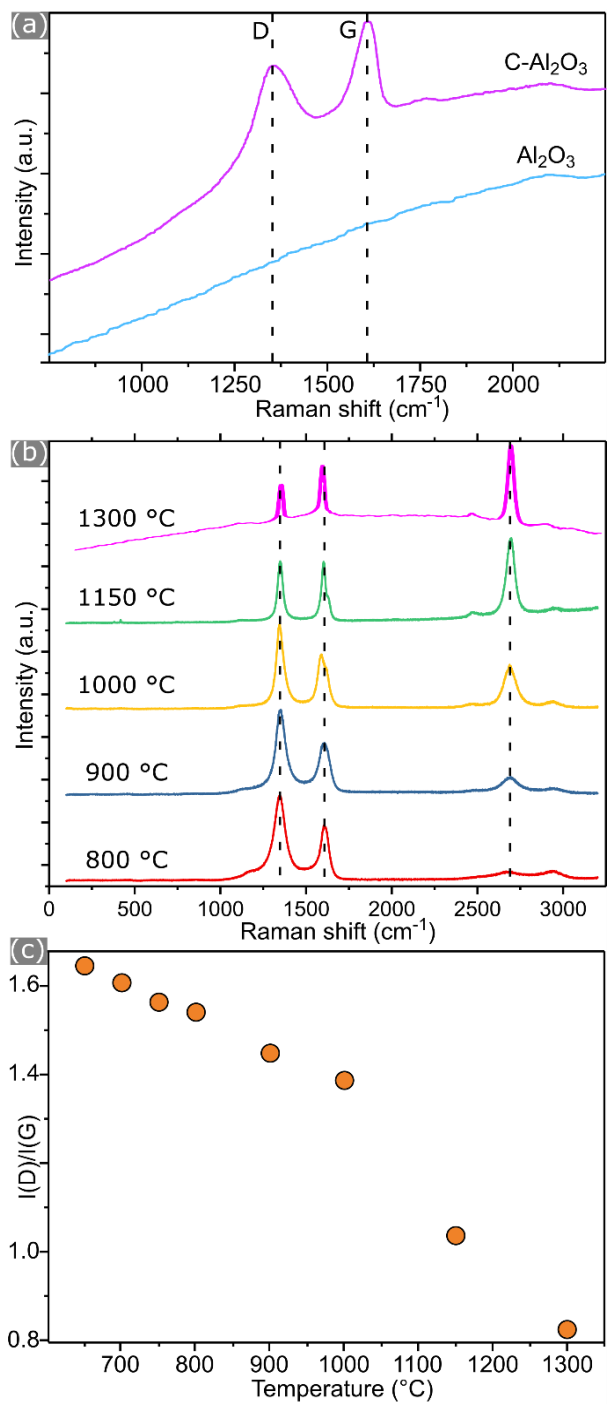

*Figure 2*. (a) Typical Raman spectra of uncoated/coated alumina nanoparticles. The Raman spectrum shows that there exists a significant amount of carbon on the surface of the nanoparticles after plasma treatment (D and G peaks are at 1347 $cm^{-1}$ and 1607 $cm^{-1}$, respectively). Pristine alumina particles did not show any detectable Raman resonance in this region. (b) The influence of the temperature on the chemistry of the grown layer. The amorphous carbon starts to graphitize at higher temperatures and forms ordered $sp^2$ structure. The intensity of the Raman peak at 2691 $cm^{-1}$ (second order resonance of the $sp^2$ network) increases with temperature. (c) The intensity ratio between D and G peaks (I(D)/I(G)) with temperature. The ratio decreases steadily with temperature, suggesting that the chemistry of the grown film is temperature dependent. At higher temperatures the coating contains more ordered $sp^2$ bonded carbon.

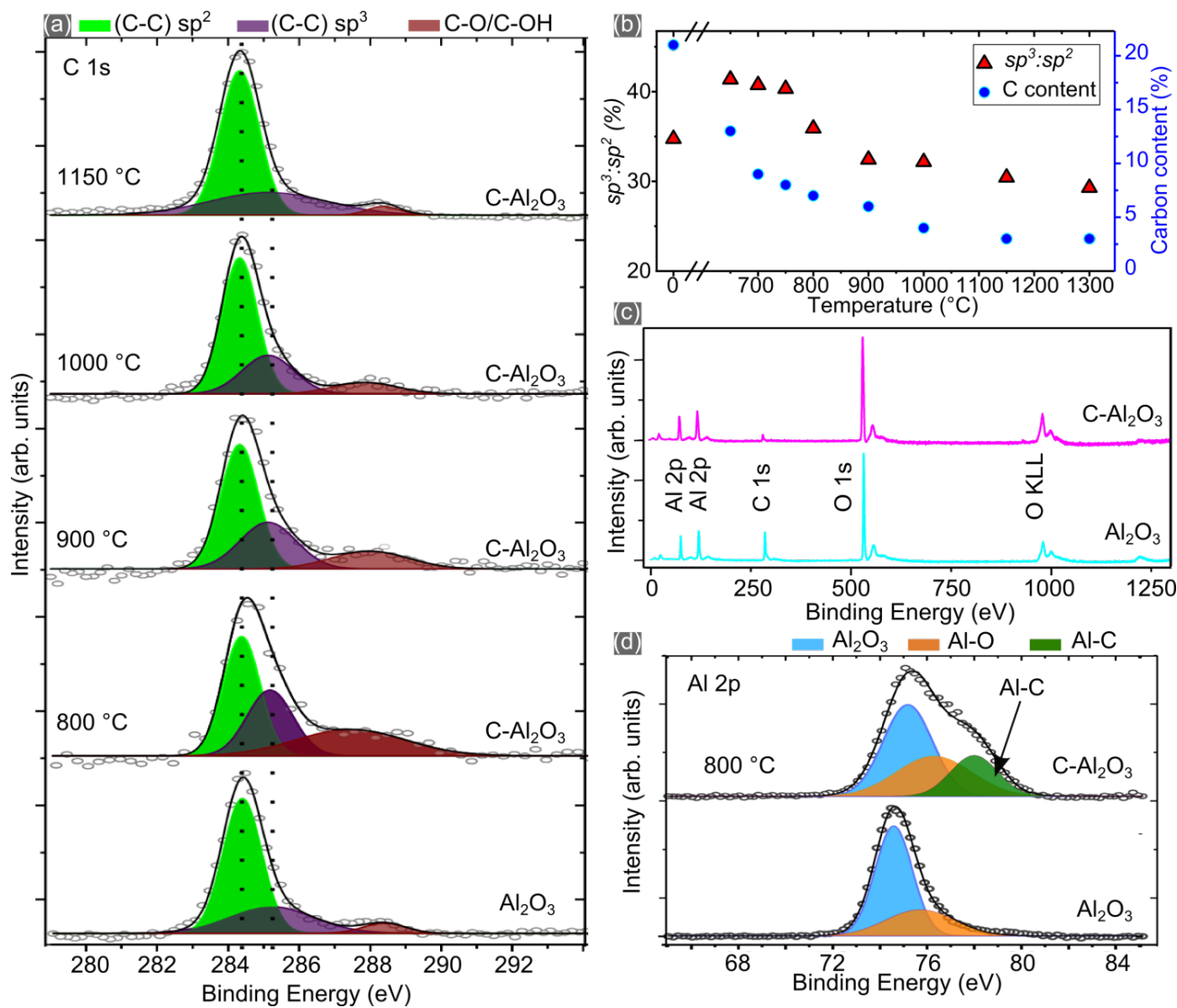

*Figure 3*. XPS spectra (E = 1253.6 eV) of uncoated/coated alumina nanoparticles. (a) Normalized C1s spectrum of the samples grown at different plasma temperatures. The $sp^2$ and $sp^3$ carbon content is temperature dependent. (b) The ratio of the $sp^3$ and $sp^2$ carbon, as a function of temperature. $sp^3$ content decreases with temperature (data obtained from high-resolution C 1s spectrum). The total carbon content of the particles also reduces significantly with temperature (data obtained by comparing the relative XPS survey spectrum). (c) The XPS survey spectrum shows that the carbon content of the coated particles is lower compared to the uncoated particles. (d) Al 2p spectra show the presence of Al-C compositions in the coated samples.

The feasibility of using these particles for *in vitro* biosensing was tested, in which the concept is demonstrated in **Figure 4**. In this approach densely-packed DNAs can be attached to the surface of nanoparticles by physisorption (*e.g.* by π-π stacking or electrostatic forces) [12]. Upon adsorption of the fluorescently-tagged DNA molecules to the surface of the particles, the fluorescence of DNA can be quenched. This type of quenching is known as "contact quenching" in the literature and it is most probably due to Förster resonance mechanism, in which the local density of states near the surface can induce changes in the electronic structure of the molecular dyes in close proximity [13]. The quenched fluorescence can be recovered after desorption of the molecule from the surface of the particle.

Competitive forces – such as induced forces by a complementary DNA and hybridization effect – may cause desorption of DNA from the surface [14]. In other words, the difference between the binding affinities of a single-stranded and double-stranded DNAs with the surface of the nanoparticle is the key factor in this competitive-assay. Due to the selection requirement, up to now the number of the materials which have been successfully used in this assay is very limited. Carbon nanomaterials – due to their surface properties – are amongst the most promising platforms for this competitive assay [15]. The good performance of carbon materials in contact quenching is attributed to the strong π-π stacking of a single-stranded DNA and $sp^2$-bonded carbon atoms on the surface of the material, which brings the nucleobases to the close proximity of the surface (~ 3.5 Å) where the van der Waals attractive forces could improve adsorption and allow the nucleobases to make direct and stable contacts with the surface without any hydration layer at the interface [16]. The presence of complimentary DNA and hybridization forces cause slight separation of the DNA molecules from the surface and subsequently the replaced hydration layer at the interface would not allow stable adsorption of the double-stranded DNA at room temperature [16].

Carbon-coated alumina particles that were grown at 800 ºC with 2000 W plasma power (in 1% $CH_4$ in $H_2$) were tested for this assay. The selection was made due to relatively high $sp^2$-bonded carbon content achieved at 800 ºC and also due to the fact that the intrinsic fluorescence of the particles was less influenced at this temperature (fluorescence was reduced at higher temperatures due to thermal annealing of the crystal defects/color centers). The average size of the particles was ~50 nm after colloidal solution preparation (see SI). Despite the ultra-small thickness of the nanocarbon coating, water dispersion of the coated particles was significantly improved, most probably due to the formation of new surface groups (such as C-OH and C-H) and the subsequent increase in the ζ-potential of the particles (please refer to SI). It is worth mentioning that alumina nanoparticles (uncoated) are not expected to perform well in this assay, due to two limiting factors: (i) limited water dispersion and (ii) limited quenching efficiency (SI).

**Figure 4(a)** shows the fluorescence spectrum of a 5 nM solution of tagged-DNA. The fluorescent emission was decreased by addition of coated nanoparticles at different concentrations. Addition of 20 µL of nanoparticle solution with the concentration of 50 µg mL$^{-1}$ completely quenched the fluorescence of 980 µL solution (the detection limit of the fluorimeter was 0.1 nM). **Figure 4(b)** shows that the quenching efficiency ($F_0/F$) was enhanced by increasing the number of nanoparticles ($F_0$ and $F$ are fluorescent intensity before and after nanoparticle addition, respectively). Enhancement of quenching efficiency by number of particles is most probably due to the fact that larger surface area provides more adsorption sites and hence leads to a higher probability for fluorescent quenching.

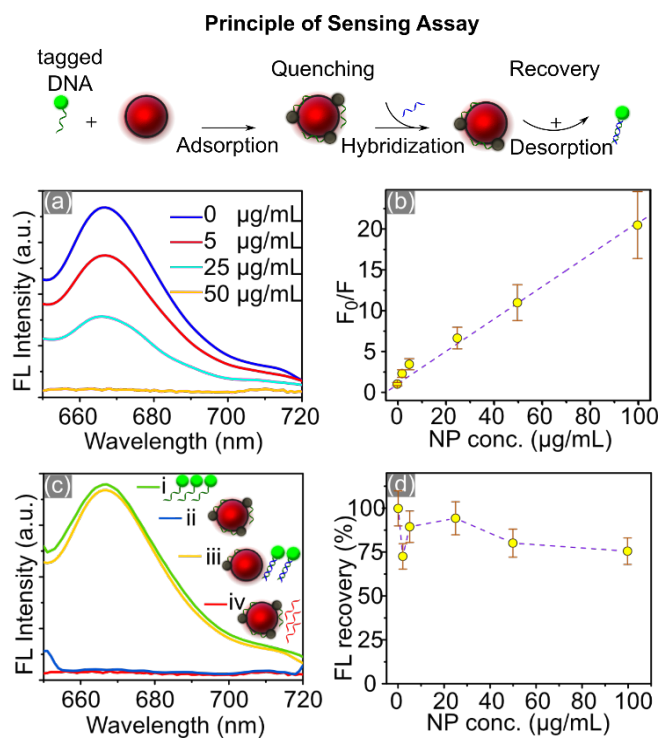

*Figure 4*. Proposed principles of competitive assay with carbon-coated nanoparticles. (a) Fluorescent intensity of a solution containing tagged DNA is reduced by addition of the carbon-coated particles. (b) The effect of nanoparticle concentration on quenching efficiency. Addition of nanoparticles enhances the quenching efficiency. (c) Fluorescence emission spectra of: (i) probe DNA (5 nM); (ii) 5 nM probe DNA + CANP; (iii) 5 nM probe DNA + CANP + non-complementary DNA; (iv) 5 nM probe DNA (excitation wavelength was 630 nm). (d) The effect of nanoparticle concentration on fluorescence recovery.

Curve (i) in **Figure 4(c)** shows the emission spectrum from a 5 nM solution of probe DNA before the addition of nanoparticles (the excitation wavelength is centered at 630 nm). Addition of coated particles into the solution reduced the fluorescence intensity (up to ~ 100%), indicating that the particles adsorbed DNA molecules and effectively quenched the attached fluorescent dye, as evidenced by curve (ii). Upon mixing with target molecules (complementary DNA) over a 20 min period, the solution recovered its fluorescence, leading to more than 90% fluorescence recovery, see curve (iii). The selectivity of the sensing platform was tested by using a non-complementary DNA as a control. The data plotted in curve (iv) suggest that the fluorescence of the solution did not have any obvious change upon addition of non-complementary DNA. These results indicate that the fluorescence recovery is due to the hybridization of DNA molecules and subsequent desorption from the surface of nanoparticles. Photoluminescence of alumina is absent in the shown spectra because there is no absorption/emission line for alumina at this excitation wavelength (630 nm).

Fluorescence recovery ($F_{recovered}/F_0$) compares the emission intensity before ($F_0$) and after addition of complementary DNA into the quenched conjugates of nanoparticle-probe DNA ($F_{recovered}$). **Figure 4(d)** shows that the fluorescence recovery was slightly reduced at higher nanoparticle concentrations, which is most probably due to the increased surface area of the nanoparticles. In this study adsorption of tagged DNA was investigated by fluorescence measurements. However, alternative techniques such as polyacrylamide gel electrophoresis (PAGE) may allow quantitative measurements of the adsorbed/desorbed molecules to cross-reference the results based on the analysis of fluorescence.

In addition to solution-based DNA sensing, we explored the proof-of-concept possibility of using carbon-coated alumina particles for live cell imaging. First, internalization of the particles was tested by incubation of cells with coated particles. The incubated cells were fixed and stained for optical imaging. Due to the strong scattering of the particles under white-light illumination, it was possible to localize the particles at different confocal depths (please refer to SI). The results show that after incubation, the 50 nm particles were able to penetrate through the cell membrane and reach to the cell body and spread through the branches (**Figure S2-S5**).

Due to the cytocompatibility of the coated particles (SI), it was possible to locate and track the particles inside the cells using their intrinsic photoluminescence (methods and **Figure S3-S5**). Indeed, multi-functionality of the coated particle potentially allows simultaneous cell imaging and biosensing. **Figure 5** shows that DNA-loaded particles can release their payload inside a living cell which contains complementary DNA. For this purpose the particles were loaded with fluorescently tagged DNA (AF488). The intrinsic nanoparticles emission (~690-710 nm) was different from the used molecular dye (~510 nm). The cells which contained the complementary DNA exhibited green fluorescent emission after addition of the nanoparticles, while no significant fluorescence was observed for the cells without complementary elements. Excessive entrapment of the nanoparticles inside the cell is detrimental for their release efficiency. At higher particle concentrations, it is very likely that the particles form aggregates inside the living cells (**Figure S4**), reducing their efficiency to release their payloads (**Figure 5**).

These results suggest that carbon-coated alumina nanoparticles are potentially able to selectively deliver and release gene/DNA to living cells (with specific markers), without any harmful effects on normal non-specific cells (with no markers inside). Thus, the selective release and fluorescent-quenching capability of the functionalized nanocarbon combined with the intrinsic optical properties of the core alumina can open up intriguing avenues for further development of gene therapy studies and drug discovery.

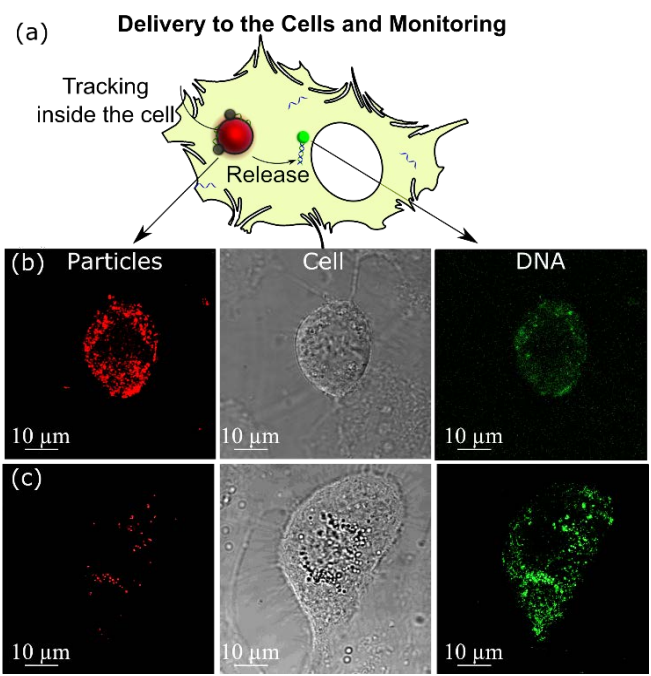

*Figure 5*. *Schematic (not to scale) of the platform for simultaneous cellular monitoring and targeted gene release. DNA loaded nanoparticles are internalized by the cell, and the DNA is released from the surface of the particle upon hybridization with its complementary pair. (b,c) show single cells used for in situ delivery and detection. The intrinsic fluorescence of the nanoparticles was detected in the red-channel (left column). The presence of the target DNA inside the living cell could be detected in green channel (right column) when they interact with the internalized nanoparticles. The middle column shows the bright field images of the living single cells. The concentration of the particles was 1000 times more in (b) compared to (c). However, more DNAs were released from the particles in (c) compared to (b). No significant release was observed in the cells without complementary pairs.*

In summary, a new type of hybrid nanocarbon/metal-oxide nanostructure was introduced via a simple and scalable fabrication process. Conformal nanocarbon coatings were produced on the surface of metal-oxide nanoparticles, using a plasma-based fabrication technique. The ultra-thin carbon layer on the surface of the alumina nanocrystals gives new surface properties to the core metal-oxide structure, such as cytocompatibility, water dispersibility, and tuneable surface chemistry, while the core particle maintains some of its key properties, such as excellent photoluminescent properties. A proof-of-principle demonstration showed that the coated particles can be used as effective carriers for DNA sensing. The sensing mechanism was based on a hybridization competitive assay: the quenched fluorescence of physisorbed DNA could be recovered upon hybridization and subsequent desorption. We provided another proof-of-principle study for using the coated particles for simultaneous live cell imaging and intra-cellular DNA sensing. This study opens up a new avenue for further systematic studies on *in vitro* gene delivery and drug discovery.

**Acknowledgements**
Authors acknowledge research facility and technical assistance from the CARF (Central Analytical Research Facility at Queensland University of Technology). Authors thank the technical assistance from Dr. Leonore Deboer and access to the IHBI facility at QUT for cell imaging. MA acknowledges Marie Skłodowska-Curie actions (Project Reference: 706930).